\documentclass[12pt]{iopart}
\usepackage{iopams}
\usepackage{graphicx}
\usepackage{color}
\usepackage{amsfonts}
\input amssym.def 
\input amssym

\begin{document}

\title[]{Zoology of Atlas-groups: dessins d'enfants, finite geometries and quantum commutation}

\author{ Michel Planat$^1$ and Hishamuddin Zainuddin$^2$
}

\vspace*{.1cm}
\address
{
$^1$
 Institut FEMTO-ST, CNRS, 15 B Avenue des Montboucons, F-25033 Besan\c con, France ({\tt michel.planat@femto-st.fr}),

$^2$Laboratory of Computational Sciences and Mathematical Physics, Institute for Mathematical Research, Universiti Putra Malaysia, 43400 UPM Serdang, Selangor,
Malaysia ({\tt hisham@upm.edu.my}).}

\vspace*{.2cm}

\begin{abstract}

Every finite simple group $P$ can be generated by two of its elements. Pairs of generators for $P$ are available in the Atlas of finite group representations as (not neccessarily minimal) permutation representations $\mathcal{P}$. It is unusual but significant to recognize that a $\mathcal{P}$ is a Grothendieck's {\it dessin d'enfant} $\mathcal{D}$ and that most standard graphs and finite geometries $\mathcal{G}$ - such as near polygons and their generalizations - are stabilized by a $\mathcal{D}$. In our paper, tripods $\mathcal{P}-\mathcal{D}-\mathcal{G}$ of rank larger than two, corresponding to simple groups, are organized into classes, e.g. symplectic, unitary, sporadic, etc (as in the Atlas). An exhaustive search and characterization of non-trivial point-line configurations defined from small index representations of simple groups is performed, with the goal to recognize their quantum physical significance. All the defined geometries $\mathcal{G}'s$ have a contextuality parameter close to its maximal value $1$.
\vspace*{.2cm}


\end{abstract}

\section{Introduction}
\noindent

Over the last years, it has been recognized that the detailed investigation of commutation between the elements of generalized Pauli groups -the qudits and arbitrary collections of them \cite{Pauli2011}- is useful for a better understanding of concepts of quantum information such as error correction \cite{Terhal2015,Planat2010}, entanglement \cite{Jozsa2000, Holweck2015} and contextuality \cite{Howard2014, Winter2014}, that are cornerstones of quantum algorithms and quantum computation. Only recently the first author observed that much of the information needed is encapsulated in permutation representations, of rank larger than two, available in the Atlas of finite group representations \cite{Atlasv3}. The coset enumeration methodology of the Atlas was used by us for deriving many finite geometries underlying quantum commutation and the related contextuality \cite{Dessins2014}-\cite{Moonshine2015}. As a bonus, the two-generator permutation groups and their underlying geometries may luckily be considered as {\it dessins d'enfants} \cite{Lando2004}, although this topological and algebraic aspect of the finite simple (or not simple) groups is barely mentioned in the literature. Ultimately, it may be that the Monster group and its structure fits our quantum world, as in Dyson's words \cite{Moonshine2015}. More cautiously, in Sec. \ref{Part1} of the present paper, we briefly account for the group concepts involved in our approach by defining a tripod $\mathcal{P}-\mathcal{D}-\mathcal{G}$. One leg $\mathcal{P}$ is a desired two-generator  permutation representation of a finite group $P$ \cite{Atlasv3}. Another leg $\mathcal{D}$ signs the coset structure of the used subgroup $H$ of the two-generator free group $G$ (or of a subgroup $G'$ of $G$ with relations), whose finite index $[G,H]=n$ is the number edges of $\mathcal{D}$, and at the same time the size of the set on which $\mathcal{P}$ acts, as in \cite{Planat2015}. Finally, $\mathcal{G}$ is the geometry with $n$ vertices that is defined/stabilized by $\mathcal{D}$ \cite{Dessins2014}. Then, in Sec. \ref{Part2}, we organize the relevant $\mathcal{P}-\mathcal{D}-\mathcal{G}$ tripods taken from the classes of the Atlas and find that many of them reflect quantum commutation, specifically the symplectic, unitary and orthogonal classes. The geometries of other (classical and sporadic) classes are investigated similarly with the goal to recognize their possible physical significance. A physically oriented survey of simple groups is \cite{Boya2013}.

\section{Group concepts for the $\mathcal{P}-\mathcal{D}-\mathcal{G}$ puzzle}
\label{Part1}

\subsection{Groups, dessins and finite geometries}
\noindent

Following the impetus given by Grothendieck \cite{Groth84}, it is now known that there are various ways to dress a group $P$ generated by two permutations,
(i) as a connected graph drawn on a compact oriented two-dimensional surface -a bicolored map (or hypermap) with $n$ edges, $B$ black points, $W$ white points, $F$ faces, genus $g$ and Euler characteristic $2-2g=B+W+F-n$ \cite{Jones78}, (ii) as a Riemann surface $X$ of the same genus equipped with a meromorphic function $f$ from $X$ to the Riemann sphere $\bar{\mathbb{C}}$ unramified outside the critical set $\{0,1,\infty\}$ -the pair $(X,f)$ called a Belyi pair and, in this context, hypermaps are called {\it dessins d'enfants}  \cite{Lando2004,Groth84},
 (iii) as a subgroup $H$ of the free group $G=\left\langle a,b\right\rangle$ where $P$ encodes the action of (right) cosets of $H$ on the two generators $a$ and $b$ -the Coxeter-Todd algorithm does the job \cite{Planat2015} and finally (iv), when $P$ is of rank at least three, that is of point stabilizer with at least three orbits, as a non-trivial finite geometry \cite{Dessins2014}-\cite{Moonshine2015}. Finite simple groups are generated by two of their elements \cite{Malle1994} so that it is useful to characterize them as members of the categories just described.

There are many mathematical papers featuring the correspondence between items (i) and (ii) in view of a better understanding of the action of the {\it absolute Galois group} $\mbox{Gal}(\bar{\mathbb{Q}}/\mathbb{Q})$ -the automorphism group of the field $\bar{\mathbb{Q}}$ of algebraic numbers- on the hypermaps \cite{Groth84,Jones78,Girondo}. Coset enumeration featured in item (iii) is at work in the permutation representations of finite groups found in the Atlas \cite{Atlasv3}. Item (i) in conjunction to (iii) and (iv) allowed us to arrive at the concept of geometric contextuality as a lack of commutativity of cosets on the lines of the finite geometry stabilized by $P$ \cite{Planat2015}.

Item (iv) may be further clarified thanks to the concept of rank of a permutation group $P$. First it is expected that $P$ acts faithfully and transitively on the set $\Omega=\{1,2,\cdots, n\}$ as a subgroup of the symmetric group $S_n$. The action of $P$ on a pair of distinct elements of $\Omega$ is defined as $(\alpha,\beta)^p=(\alpha^p,\beta^p)$, $p\in P$, $\alpha \ne \beta$. The orbits of $P$ on $\Omega \times \Omega$ are called orbitals and the number of orbits is called the rank $r$ of $P$ on $\Omega$. The rank of $P$ is at least two and the $2$-transitive groups identify to the rank $2$ permutation groups. Second the orbitals for $P$ are in one to one correspondence with the orbits of the stabilizer subgroup $P_{\alpha}=\{p \in P |\alpha^p=\alpha\}$ of a point $\alpha$ of $\Omega$. It means that $r$ is also defined as the number of orbits of $P_{\alpha}$.  The orbits of $P_ {\alpha}$ on $\Omega$ are called the suborbits of $P$ and their lengths are the subdegrees of $P$. A complete classification of permutation groups of rank at most $5$ is in the book  \cite{Praeger97}. Next, selecting a pair $(\alpha,\beta)\in \Omega \times \Omega$, $\alpha \ne \beta$, one introduces the two-point stabilizer subgroup $P_{(\alpha,\beta)}=\{p \in P|(\alpha,\beta)^p=(\alpha,\beta)\}$. There exist $1 < m \le r$ such non isomorphic (two-point stabilizer) subgroups $S_m$ of $P$. Selecting the largest one with $\alpha \ne \beta$, one defines a point/line incidence geometry $\mathcal{G}$ whose points are the elements of $\Omega$ and whose lines are defined by the subsets of $\Omega$ sharing the same two-point stabilizer subgroup. Thus, two lines of $\mathcal{G}$ are distinguished by their (isomorphic) stabilizers acting on distinct subsets of $\Omega$. A non-trivial geometry arises from $P$ as soon as the rank of the representation $\mathcal{P}$ of $P$ is $r>2$ and simultaneously the number of non isomorphic two-point stabilizers of $\mathcal{P}$ is $m>2$. 

\subsection{Geometric contextuality}
\noindent

Let $G'$ be a subgroup of the free group $G=\left\langle a,b\right\rangle$ endowed with a set of relations and $H$ a subgroup of $G$ of index $n$. As shown in Sec. 2.1, the permutation representation $\mathcal{P}$ associated to the pair $(G',H)$ is a dessin d'enfant $\mathcal{D}$ whose edges are encoded by the representative of cosets of $H$ in $G'$. A graph/geometry $\mathcal{G}$ may be defined by taking the $n$ vertices of $\mathcal{G}$ as the edges of $\mathcal{D}$ and the edges of $\mathcal{G}$ as the distinct (but isomorphic) two-point stabilizer subgroups of $\mathcal{P}$.

Further, $\mathcal{G}$ is said to be {\it contextual} if at least one of its lines/edges corresponds to a set/pair of vertices encoded by non-commuting cosets \cite{Planat2015}. A straightforward measure of contextuality is the ratio $\kappa=E_c/E$ between the number $E_c$ of lines/edges of $\mathcal{G}$ with non-commuting cosets and the whole number $E$ of lines/edges of $\mathcal{G}$. Of course, lines/edges passing through the identity coset $e$ have commuting vertices so that one always as $\kappa<1$.

In Sec. 3 below, the contextuality parameter $\kappa$ corresponding to the collinear graph of the relevant geometry $\mathcal{G}$ is displayed in the right column of the tables. In order to compute $\kappa$, one needs the finite presentation of the corresponding subgroup $H$ in $G'$ leading to the permutation representation $\mathcal{P}$ but this information is not always available in the Atlas.

\subsection{A few significant geometries}
\noindent

There exist layers in the organization of finite geometries, see \cite{Batten1997} for an introduction. A partial linear space is an incidence structure $\Gamma(P,L)$ of points $P$ and lines $L$ satisfying axioms (i) any line is at least with two points and (ii) any pair of distinct points is incident with at most one line. In our context, the geometry $\mathcal{G}$ that is defined by a two-generator permutation group $\mathcal{P}$, alias its dessin d'enfant $\mathcal{D}$, has order $(s,t)$ meaning that every line has $s+1$ points and every point is on $t+1$ lines. Thus $\mathcal{G}$ is the geometric configuration $[p_{s+1},l_{t+1}]_{(r)}$, with $p$ and $l$ the number of points and lines. The extra index $r$ denotes the rank of $\mathcal{P}$ from which $\mathcal{D}$ arises.

We introduce a first layer of organization that is less restrictive that of a {\it near polygon} to be defined below and that of a {\it symplectic polar space} encountered in Sec. 3.3. We denote by $\mathcal{G}_u=\mathcal{G}(s,t;u)$ a connected partial linear space with the property that, given a line $L$ and a point $x$ not on $L$, there exist
 a constant number $u$ of points
 of $L$ nearest to $x$. A {\it near polygon} (or near $2d$-gon) is a partial linear space such that the maximum distance between two points (the so-called diameter) is $d$ and, given a line $L$ and a point $x$ not on $L$, there exists \lq a unique point' on $L$ that is nearest to $x$. A graph (whose lines are edges) is of course of type $\mathcal{G}_1$. 
 A near polygon is, by definition, of type $\mathcal{G}_1$.
Symplectic polar spaces are of the form $\mathcal{G}_u$, possibly with $u>1$, but not all $\mathcal{G}_u$ with $u>1$ are polar spaces. A {\it generalized polygon} (or generalized $N$-gon) is a near polygon whose incidence graph has diameter $d$ (the distance between its furthest points) and girth $2d$ (the length of a shortest path from a vertex to itself). According to Feit-Higman theorem \cite{Tits2002}, finite generalized $N$-gons with $s>1$ and $t>1$ may exist only for $N \in \{2,3,4,6,8\}$. They consist of projective planes with $N=3$, and generalized quadrangles GQ$(s,t)$, generalized hexagons GH$(s,t)$ and generalized octagons GO$(s,t)$ when $N=4,6,8$, respectively.

Many $\mathcal{G}'s$ have a collinearity graph that is a strongly regular graph (denoted srg). These graphs are partial geometries pg$(s,t;\alpha)$ of order $(s,t)$ and (constant) connection number $\alpha$. By definition, $\alpha$ is the number of points of a line $L$ joined to a selected point $P$ by a line. The partial geometries pg listed in our tables are those associated to srg graphs found in \cite{Brouwer}.

\subsection{A few small examples}
\noindent

Let us illustrate our concepts by selecting a rank $3$ (or higher) representation for the group of the smallest cardinality in each class of simple groups. The notation for the simple groups and their representations are taken from the Atlas \cite{Atlasv3}.

\subsubsection*{Alternating}
\noindent

The smallest non-cyclic simple group is the alternating group $A_5$ whose finite representation is
$H=\left\langle a,b|a^2=b^3=(ab)^5=1\right\rangle$.

The permutation representations of $A_5$ are obtained by taking the subgroups of finite index of the free group $G=\left\langle a,b\right\rangle$ whose representation is $H$. 

\begin{table}[ht]
\begin{center}
\begin{tabular}{|l|crcrc|}
\hline \hline
$A_5$ index & $5$ & $6$ & $10$ & $12$ &$15$\\
\hline
$r$ & $2$ &  $2$ & $3$  &  $4$ & $5$\\
$m$ & $2$ &  $2$ & $3$  &  $1$ & $1$\\
\hline
\hline
\end{tabular}
\caption{Parameter $r$ and $s$ for small index representations of $A_5$.}
\end{center}
\end{table}

Table 1 list the rank $r$ and the number $m$ of two-point stabilizer subgroups for the  permutation representations $\mathcal{P}$ up to rank $15$. The only non trivial permutation group has index $10$, rank $3$, subdegrees $1,3,6$ with 
$\mathcal{P}=\left\langle 10|(2,3,4)(5,7,8)(6,9,10), (1,2)(3,5)(4,6)(7,10) \right\rangle$.

The dessin d'enfant $\mathcal{D}$ corresponding to $\mathcal{P}$ is pictured in our previous papers, see \cite[Fig. 10]{Dessins2014}, \cite[Fig. 3j]{Planat2015}, \cite[Fig.  4]{Moonshine2015}. The geometries that are stabilized are the Petersen graph PG, or Mermin's pentagram MP, depending on the choice of the two-point stabilizer subgroup. Thus $A_5$ features three-qubit \lq 3QB' contextuality.

\subsubsection*{Symplectic}
\noindent

The smallest (simple) symplectic group is $S'_4(2)=A_6$ whose finite representation is
$H=\left\langle a,b|a^2=b^4=(ab)^5=(ab^2)^5=1\right\rangle$.
Table 2 list the rank $r$ and the number $m$ of two-point stabilizer subgroups for the  permutation representations $\mathcal{P}$ up to rank $30$.

The smallest non trivial permutation group $\mathcal{P}$ has index $15$, rank $3$ and subdegrees $1,6,8$ as shown in Table 2.

\begin{table}[ht]
\begin{center}
\begin{tabular}{|l|crcrc|}
\hline \hline
$A_6$ index & $6$ & $10$ & $15$ & $20$ &$30$\\
\hline
$r$ & $2$ &  $2$ & $3$  &  $4$ & $7$\\
$m$ & $2$ &  $2$ & $3$  &  $2$ & $3$\\
\hline
\hline
\end{tabular}
\caption{Parameter $r$ and $s$ for the small index representations of $A_6$.}
\end{center}
\end{table}

The geometry that is stabilized by $\mathcal{P}$ is the (self-dual) generalized quadrangle $GQ(2,2)$, alias the graph $\hat{L}(K_6)$ (the complement of line graph of the complete graph $K_6$). It is known that $GQ(2,2)$ is a model of two-qubit \lq 2QB' commutation, see \cite[Fig. 12]{Planat2015}. 
The permutation representation of index $30$ of $S'_4(2)$ stabilizes the configuration $[30_{16},160_3]$ of rank $7$ that turns to be a geometry of type $\mathcal{G}_2$.

\begin{figure}[ht]
\centering 
\includegraphics[height=8cm]{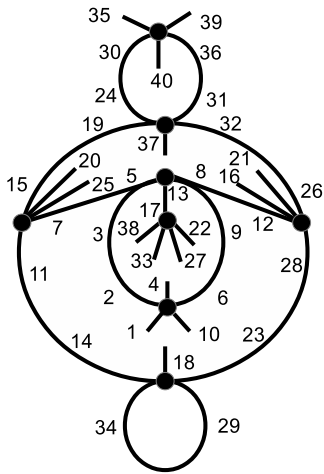}
\caption{The dessin d'enfant stabilizing the generalized quadrangle $GQ(3,3)$ (a model of two-qutrit \lq 2QT' commutation). The dessin corresponds to the Sp$(4,3)$ permutation representation of index $40_b$ found in the Atlas. Only black points are shown: white points are implicit at the mid-edges or at the ends of half-edges.}
\label{dessin40}
\end{figure}

As for two-qutrit commutation, one uses the $S_4(3)$ permutation representation $\mathcal{P}$ of rank $3$ and index $40_b$ found in the Atlas. The dessin d'enfant picturing $\mathcal{P}$ is found on Fig. \ref{dessin40}. The dessin has signature $(B,W,F,g)=(8,28,6,0)$.

\subsubsection*{Unitary}
\noindent

The smallest (simple) unitary group is $U_3(3)$. Representations of $U_3(3)$ of index $28$ (rank $2$), $36$ (rank $4$), $63$ (rank $4$) and $63$ (rank $5$) may be found in the Atlas (denoted $63_a$ and $63_b$, respectively). The most interesting ones are the $63_a$, of subdegrees $1,6,24,32$ and the $63_b$, of subdegrees $1,6,16^2,24$. These representations stabilize the split Cayley hexagon $GH(2,2)$ (with $63_b$) and its dual  (with $63_a$). The hexagon $GH(2,2)$ is a configuration of type $[63_3]$ with $63$ points on three lines and $63$ lines with three points. It may be used as a model of $3QB$ contextuality, see \cite[Fig. 5 and 6]{Planat2015} for details and plots of the corresponding dessins d'enfants.

\subsubsection*{Orthogonal}
\noindent

The smallest (simple) orthogonal group is $O_7(3)$. The Atlas lists four representations of rank $3$ and index $351$, $364$, $378$ and $1080$. We could recognize that the first representation is associated to the strongly regular graph srg($351,126,45,45$) and the geometry $NO^-(7,3)$,
the second representation is associated with srg($364,120,38,40$) and the geometry of the symplectic polar space $W_5(3)$, the third representation is associated with srg($378,117,36,36$) and presumably the partial geometry pg$(13,18,4)$, and the fourth representation is associated with srg($1080,351,126,108$) and the geometry $NO^+(8,3)$, see \cite{Brouwer}  for details about the undefined acronyms. The second representation corresponds to the commutation of the $364$ three-qutrit \lq 3QT' observables \cite{Pauli2011}. It is found to be of type $\mathcal{G}_4$. The representation of index $1120$ and rank $4$ of $O_7(3)$ found in the Atlas is associated to the dual of $W_5(3)$ that is the dense near hexagon $DQ(6,3)$. See table 9 for further details.

\footnotesize
\begin{table}[ht]
\begin{center}
\begin{tabular}{|l|c|r|c|r|c|r|}
\hline \hline
type&  group & $m$& $r$ & $\mathcal{G}$ & physics & $\kappa$\\
\hline\hline
alternating & $A_5$ &  $10$ & $3$  &  PG $=\hat{L}(K_5)$ & MP in 3QB & 0.767\\
linear & $L_2(5)=A_5$ & .& .  &  .& . & ;\\
symplectic & $S'_4(2)$ & $15$& $3$ & $GQ(2,2) ={\hat{L}(K_6)}$ & $2$QB & .\\
. & $S_4(3)$ & $40$ & $3$ & $GQ(3,3)$  & $2$QT & 0.800\\
unitary & $U_3(3)$ &  $63$ & $4$ &  $GH(2,2)$& $3$QB & 0.704\\
orthogonal & $O_7(3)$ &  $364$& $3$ &  $W_5(3)$, $\mathcal{G}_4$ & $3$QT & 0.846\\
except. untwist. & $G_2(2)'=U_3(3)$&. &  .  &  .& . & .\\
except. twist. & Sz$(8)$ &  $560$  & $17$& $[560_{13},1820_4]_{(17)}$& ? & 0.971\\
sporadic & $M_{11}$ &  $55$  &  $3$ &  $T(11)=L(K_{11})$& ? &\\
\hline
\hline
\end{tabular}
\caption{A few characteristics of a index $m$ and rank $r=3$ (or higher) representation of the simple group of smallest cardinality in each class. The characteristics of the Sp$(4,3)$ representation for two qutrits is added to this list. The question marks point out that a physical interpretation is lacking.}
\end{center}
\end{table}
\normalsize

\subsubsection*{Exceptional and twisted}
\noindent

The smallest (simple) twisted exceptional group is Sz$(8)$. The representation of index $520$ listed in the Atlas leads to an unconnected graph. The representation of index $560$ of rank $17$ and subdegrees $1,13^3,26^6,52^7$ leads to a configuration of type $[560_{13},1820_4]$ (i.e. every point is on $13$ lines and there are $1820$ lines of size $4$). The Atlas also provides a representation of index $1456$ and rank $79$ that leads to another geometry, of order $(3,4)$, with again $1820$ lines of size $4$ (see also the relevant item in table 10). The physical meaning of both representations, if any, has not been discovered.

\subsection*{Sporadic}
\noindent

The smallest sporadic group is $M_{11}$. The Atlas provides representations of rank $3$ and index $55$, rank $4$ and index $66$, and rank $8$ and index $165$. The first representation leads to the triangular graph $T(11)=L(K_{11})$. The second one leads two a non strongly regular graph with $495$ edges, of girth $4$ and diameter $2$.
The third representation leads to a partial linear space of order $(2,3)$ with $220$ lines/triangles.

\subsubsection*{Brief summary}
\noindent

The results of this subsection are summarized in Table 3. Observe that the smallest simple linear group is equivalent to $A_5$ and that the smallest untwisted group $G_2(2)'$ is similar to $U_3(3)$. Except for $M_{11}$ and Sz$(8)$ all these \lq small' groups occur in the commutation of  quantum observables. Further relations between the geometry of simple groups and the commutation of multiple qudits are given at the next section.

\section{Atlas classes and the related geometries}
\label{Part2}
\noindent

\subsection{Alternating}
\noindent

The non trivial configurations that are stabilized by (low rank) small simple alternating groups are listed in Table 4. The alternating group $A_7$ is missing because no non-trivial geometry has been recognized. Permutation groups for alternating groups $A_n$, $n>8$ are those listed in the Atlas.

\begin{table}[ht]
\begin{center}
\begin{tabular}{|l|c|l|}
\hline \hline
 $A_n$ & geometries $_{(\mbox{rank})}$ ~~~~~~~~~~~   $\mathcal{G}_{(r)}$ & $\kappa$\\
\hline
$A_5$ & MP, $\hat{T}(5)_{(3)}$ & 0.767, 0.666\\
$A_6$ & $GQ(2,2)$, $\hat{T}(6)_{(3)}$ & 0.800\\
$A_8$ & $T(8)_{(3)}$, ($[35_6,30_7]_{(3)}$: srg, $\mathcal{G}_3$, $S(2,3,15)$, lines in $PG(3,2)$, $O^+(6,2)$) & 0.684, 0.737\\
$A_9$ & $T(9)_{(3)}$, $[126_5,315_2]_{(5)}$, $[280_3,84_{10}]_{(5)}$, $[840_4,1120_3]_{(12)}$ &\\
$A_{10}$ & $T(10)_{(3)}$, $[126_{10},210_6]_{(4)}$: srg), $[210_{15},1575_2]_{(5)}$, $[945_{10},3150_3]_{(7)}$ &\\
$A_{11}$ & $T(11)_{(3)}$,  $[165_8,330_4]_{(4)}$,  $[462_6,1386_2]_{(6)}$ &\\
$A_{12}$ & $T(12)_{(3)}$, $[220_9,495_4]_{(4)}$, $[462_{12},792_7]_{(4)}$ &\\
$A_{13}$ & $T(13)_{(3)}$ &\\
$A_{14}$ & $T(14)_{(3)}$, $[364_{11},1001_4]_{(4)}$ &\\
$A_{15}$ & $T(15)_{(3)}$, $[1365_{11},3003_5]_{(5)}$ &\\
\hline
\hline
\end{tabular}
\caption{The non trivial configurations stabilized by small simple alternating groups and their rank $r$ given as an index. The notation $T(n)=L(K_n)$ means the triangular graph and $S(2,k,v)$ means a Steiner system, that is, a $2-(v,k,1)$ design \cite{Brouwer}. The symbol srg is for a strongly regular graph. A description of the $A_8$ configuration on $35$ points is given in the text.}
\end{center}
\end{table}

\subsubsection*{The $A_8$ configuration on $35$ points}
\noindent

It has been shown at the previous section that $A_5$ and $A_6$ are associated to three-qubit contextuality (via Mermin's pentagram) and two-qubit commutation (via the generalized quadrangle of order two $GQ(2,2)$), respectively. Since $A_8$ encodes the $35$ lines in $PG(3,2)$, the corresponding configuration may be seen as a model of four-qubit contextuality, see \cite[Sec. 4]{Planat2015} for the recognition of $PG(3,2)$ as a model of a 4QB maximum commuting set and \cite{Levay2013} for an explicit reference to the $O^+(6,2)$ polarity. 

As the permutation representation is not in the Atlas, we provide a few details below. The permutation representation on $35$ points of $A_8$ is
\small
\begin{eqnarray}
&\mathcal{P}=<35|(3, 4, 6, 12, 10, 5)(7, 13, 19, 23, 15, 9)(8, 14, 21, 24, 16, 11)(17, 25,
    26)\nonumber \\
&(18, 27, 28)(20, 22, 30)(29, 33, 35, 34, 32, 31),(1, 2, 3)(4, 7, 8)(5, 9, 11)(12, 17, 18)\nonumber \\
 &(13, 20, 14)(15, 22, 16)(19, 29, 21)(23,    31, 24)(25, 32, 27)(26, 33, 28)>.\nonumber \\\nonumber
\end{eqnarray}
\normalsize
The representation is of rank $3$, with suborbit lengths $(1, 16, 18)$,  and corresponds to a dessin $\mathcal{D}$ of signature $(B,W,F,G)=(9,15,5,4))$, that is, of genus $4$, and cycles $[6^4 3^3 1^2,3^{10}1^5,7^5]$. The two-point stabilizer subgroups are of order $32$ and $36$. The group of order $36$ is isomorphic to the symmetry group
 $\mathbb{Z}_3^2 \times \mathbb{Z}_2^2$ of the Mermin square (a $3 \times 3$ grid), see \cite[Sec. 4.4]{Dessins2014}. 
The edges of the collinearity graph of the putative geometry $\mathcal{G}$ are defined as sharing the same stabilizer subgroup of order $36$, up to isomorphism, but acting on different subsets. The graph is srg of spectrum $[16^1,2^{20},-4^{14}]$ and can be found in \cite {Brouwer}. The lines of $\mathcal{G}$ are defined as the maximum cliques of the collinearity graph. In the present case, the lines do not all share the same stabilizer subgroup. One gets $\mathcal{G}=[35_8,56_5]_{(3)}$, a finite geometry of type $\mathcal{G}_2$.
The collinearity graph associated to the stabilizer subgroup of order $32$ is the complement of the collinearity graph of $\mathcal{G}$ and the corresponding geometry is $\bar{\mathcal{G}}=[35_6,30_7]_{(3)}$, a configuration of type $\mathcal{G}_3$, and a model of the $O^+(6,2)$ polarity.

\subsection{Linear}
\noindent

The non trivial configurations that are stabilized by (low rank) small simple linear groups are listed in Table 5.

As for a relation to physics, we already know that the linear group $L_2(4)=L_2(5)=A_5$ is associated to a 3QB pentagram and that $L_2(9)=A_6$ is associated to 2QB commutation. Then the group $L_5(2)$ is associated to $5QB$ contextuality through lines in $PG(4,2)$.
The other configurations in table 5 lack a physical meaning.

\begin{table}[ht]
\begin{center}
\begin{tabular}{|l|c|l|}
\hline \hline
group $L_n(m)$ & geometries $_{(\mbox{rank})}$ ~~~~~~~~~~~   $\mathcal{G}_{(r)}$ & $\kappa$\\
\hline
$L_2(4)=L_2(5)=A_5$ & MP, $\hat{T}(5)_{(3)}$ & 0.767, 0.666 \\
$L_2(7)$ & ( $[21_2,14_3]_{(6)}$: GH(2,1)),  $[28_3,21_4]_{(7)}$ & 0.857, 0.893 \\
$L_2(8)$ &   $[36_7,63_4]_{(5)}$: srg &\\
$L_2(9)=A_6$ &  $GQ(2,2)$, $(\hat{T}(6))_{(3)}$ & 0.800\\
$L_2(11)$ &   $[55_3]_{(9)}$ & \\
$L_2(19)$ &   $[37_6,171_2]_{(4)}$, $[171_5,285_3]_{(15)}$, $[190_{108},5130_{4}]_{(16)}$ &\\
$L_2(32)$ &   $\hat{T}(33)_{(17)}$: srg & 0.968 \\
\hline
$L_3(2)=L_2(7)$ & . & .\\
$L_3(3)$ &   $[144_{78},2308_4]_{(8)}$ & \\
$L_3(4)$ &   $[56_{10},280_2]_{(3)}$: srg, Sims-Gewirtz graph  & 0.911\\
\hline
$L_5(2)$ &   $[155_{7}]_{(3)}$: srg, $S(2,3,31)$, lines in $PG(4,2)$ & \\
\hline
\hline
\end{tabular}
\caption{The non trivial configurations stabilized by small simple linear groups and their rank. The configuration $[21_2,14_3]_{(6)}$ configuration corresponds to the thin generalized hexagon GO$(2,1)$ (see Fig. 6 of \cite{Planat2016}). 
}
\end{center}
\end{table}

\subsection{Symplectic}
\noindent

The symplectic class of simple groups is  a very useful one for modeling quantum commutation of multiple qudits. At the previous section, we already met groups $S'_4(2)$ and $S_4(3)$ associated to two-qubits and two-qutrits, respectively.

\subsubsection*{The group $S_4(3)$.}
\noindent

Let us go back to the group $S_4(3)$ whose finite representation is $H=\left\langle a,b|a^2=b^5=(ab)^9=[a,b]^3=[a,bab]^2=1\right\rangle$. Apart from $GQ(3,3)$ associated to two-qutrits other geometries exist for this group as shown in Table 6.

\begin{table}[ht]
\begin{center}
\begin{tabular}{|l|c|r|c|l|}
\hline \hline
$S_4(3)$ index & $\mathcal{D}$-signature & spectrum& geometry & $\kappa$\\
\hline
$27$ & $(7,15,3,2)$  & $[10^1,1^{20},-5^6]$       & $[27_5,45_3]_{(3)}$:   {\bf GQ(2,4)} & 0.785\\
$36$ & $(8,24,4,1)$  & $[15^1,3^{15},-3^{20}]$    & $[36_{15},135_4]_{(3)}$:  OA(6,3) & 0.833 \\
$40_b$ & $(8,28,6,0)$  & $[12^1,2^{24},-4^{15}]$    & $[40_4]_{(3)}$:  {\bf GQ(3,3)} & 0.704\\
$40_a$ & $(8,24,8,1)$  & $[12^1,2^{24},-4^{15}]$    & $[40_4]_{(3)}$:   GQ(3,3) dual & 0.825\\
$45$ & $(9,29,7,1)$  & $[12^1,3^{20},-3^{24}]$     & $[45_3,27_5]_{(4)}$: GQ(4,2) & 0.855\\
\hline
\hline
\end{tabular}
\caption{Characteristics of small index representations of $S_4(3)$ and their geometry. The bold notation correspond to geometries that are \lq stabilized' by the corresponding permutation representation $\mathcal{P}$. The other geometries that are only \lq defined' from the collinearity graph associated to  $\mathcal{P}$.}
\end{center}
\end{table}

 A few remarks are in order. Stricto sensu, only the generalized quadrangles $GQ(2,4)$ and $GQ(3,3)$ are \lq stabilized' by the corresponding permutation representations $\mathcal{P}$ (and dessins d'enfants $\mathcal{D}$ -their signature is given at the second column). The lines of each of the two geometries are defined as having two-stabilizer subgroups acting on the same subsets of points. In a weaker sense, the permutation representation for index $36$, $40_a$ and $45$ \lq define' the geometries $OA(6,3)$, the dual of $GQ(3,3)$ and $GQ(4,2)$ from the collinearity graph, its srg spectrum (shown at the third column) and the structure of its maximum cliques. In these last cases, not all lines of the geometry have their pair of points corresponding to the same two-stabilizer subgroup. Observe that case $40_a$ and case $40_b$ are isospectral but with a distinct $\mathcal{D}$-signature.

\subsubsection*{The group $S_6(2)$.}
\noindent

Another group of rich structure is the symplectic group $S_6(2)$ whose finite representation is 
 $H=\left\langle a,b|a^2=b^7=(ab)^9=[ab^2]^{12}=[a,b]^3,[a,b^2]^2=1\right\rangle$. The smallest non-trivial permutation representation $\mathcal{P}$ of  $S_6(2)$ stabilizes the symplectic polar space $W_5(3)$ associated to three-qubits \cite{Pauli2011}. The small permutation representations of $S_6(2)$ are shown on Table 7. The one of index $135$ is associated to the near quadrangle DQ$(6,2)$ \cite[chap. 6]{deBruyn2006}.
\small
\begin{table}[ht]
\begin{center}
\begin{tabular}{|l|c|r|c|l|}
\hline \hline
$S_6(2)$ & $\mathcal{D}$-signature & spectrum& geometry & $\kappa$\\
\hline
$63$ & $(9,47,7,1)$  & $[30^1,3^{35},-5^{27}]$       & $[63_{15},135_7]_{(3)}$: $\mathcal{G}_3$,  {\bf W$_5$(2)} & 0.787\\
$120$ & $(16,60,14,15)$  & $[56^1,8^{35},-4^{84}]$       & $[120_{28},1120_3]_{(3)}$& 0.847\\
$126$ & $(18,64,14,16)$  & $[64^1,8^{27},0^{63},-8^{33}]$       & $[126_{64},2688_3]_{(5)}$ & 0.766\\
$135$ & $(21,75,15,13)$  & $[14^1,5^{35},-1^{84},-7^{15}]$       & $[135_{7},315_3]_{(5)}$: {\bf DQ(6,2)} & 0.794\\
$240$ & $(36,120,28,29)$  & $[126^1,6^{84},0^{120},-18^{35}]$       & $[240_{17280},518400_8]_{(5)}$ & 0.894\\
$315$ & $(45,195,37,20)$  & $[18^1,9^{35},3^{84},-3^{195}]$       & $[315_{7},135_3]_{(5)}$ & 0.909\\
$336$ & $(48,216,40,17)$  & $[20^1,8^{35},2^{168},-4^{105},-8^{27}]$       & $[336_{10},1120_3]_{(5)}$ & 0.918\\
$960$ & $(138,480,114,115)$  & $[56^1,8^{385},-4^{504},-16^{70}]$       & $[960_{28},8960_3]_{(6)}$ & 0.961\\
\hline
\end{tabular}
\caption{Characteristics of small index representations of $S_6(2)$ and their geometry. The meaning of bold notation is as in Table 6.}
\end{center}
\end{table}
\normalsize

\subsubsection*{The geometry of multiple qudits.}
\noindent

We define the multiple qudit Pauli group $\mathcal{P}_q (q=p^n)$ as the $n$-fold tensor product between single $p$-dit Pauli operators with $\omega=\exp(\frac{2i\pi}{p})$ and $p$ a prime number.
Observables of $\mathcal{P}_q /\mbox{Center}(\mathcal{P}_q )$ are seen as the elements of the $2n$-dimensional vector space $V(2n,p)$ defined over the field $\mathbb{F}_p$.
The commutator $[.,.]:~ V(2n,p) \times V(2n,p) \rightarrow  \mathcal{P}_q'$ induces a non-singular alternating bilinear form on $V(2n,p)$, and simultaneously a symplectic form on the projective space $PG(2n-1,p)$ over $\mathbb{F}_p$.

The $|V(2n,q)|=p^{2n}$ observables of $\mathcal{P}_q/\mbox{Center}(\mathcal{P}_q)$ are mapped to the points of the symplectic polar space $W_{2n-1}(p)$ of cardinality 
$|W_{2n-1}(p)|=\frac{p^{2n}-1}{p-1} \equiv \sigma(p^{2n-1}),$ (where $\sigma(.)$ is the sum of divisor function of the argument)
and two elements of $[\mathcal{P}_q/\mbox{Center}(\mathcal{P}_q),\times]$ commute if and only if the corresponding points of the polar space $W_{2n-1}(p)$ are collinear \cite{Pauli2011}.

A subspace of $V(2n,p)$ is called totally isotropic if the symplectic form vanishes identically on it. The number of such totally isotropic subspaces/generators $g_e$ (of dimension $p^n-1$) is
$\Sigma(n)=\prod_{i=1}^n (1+p^i).$
A spread $s_p$ of a vector space a set of generators partitioning its points. One has $|s_p|=p^{n}+1$ and $|V(2n,p)|-1=|s_p|\times|g_e|=(p^{n}+1)\times (p^{n}-1)=p^{2n}-1$. A generator $g_e$ corresponds to a maximal commuting set and a spread $s_p$ corresponds to a maximum (and complete) set of disjoint maximal commuting sets. Two generators in a spread are mutually disjoint and the corresponding maximal commuting sets are mutually unbiased.

The symplectic polar spaces $W_{2n-1}(p)$ at work, alias the commutation structure of n $p$-dits may be constructed by taking the permutation representation of index $\sigma(p^{2n-1})$ of the symplectic (rank $3$) group $S_{2n}(p)$ available in the Atlas. The special case of two-qubits [with $S'_4(2)$], two-qutrits [with $S_4(3)]$, three qubits [with $S_6(2)$]. For the group $S_6(3)$, one finds two permutation representations of index $364$ and $1120$ that are similar to the ones of the same index found for the group $O_7(3)$ (see Sec. 2, item \lq Orthogonal' and Table 9). The representation of index $364$ corresponds to the commutation structure of three qutrits and the one of index $1120$ is the dual geometry encoding the non-intersection of the $1120$ maximum commuting sets of size $26$ built with the three-qutrit observables.

The collinearity graph of the polar space $W_{2n-1}(p)$ is a srg$(a,pb,b-2,b)$, with $a=a(n)=\sigma(p^{2n-1})$ and $b=b(n)=\sigma(p^{2n-3})$. The corresponding geometric configuration is of the form $[a(n)_{\Sigma(n-1)},\Sigma(n)_{(p^n-1)/(p-1)}]$.

\subsection{Unitary}
\noindent

\begin{table}[ht]
\begin{center}
\begin{tabular}{|l|c|r|c|l|}
\hline \hline
group & $\mathcal{D}$-signature &  geometry & $\kappa$\\
\hline
$U_3(3)$& $(8,24,6,0)$ & $[36_{28},336_3]_{(4)}$ &\\
& $(13,35,9,2)$ & $[63_3]_{(4)}$: {\bf GH(2,2)}& 0.846\\
& $(15,35,9,3)$ & $[63_3]_{(5)}$: {\bf GH(2,2) dual} &0.820\\
\hline
$U_3(4)$& $(70,112,16,6)$ & $[208_{6},416_3]_{(5)}$: {\bf conf. over $\mathbb{F}_{16}$} \cite{Brouwer2009} & 0.970\\
\hline
$U_3(5)$& $(10,20,8,7)$ & $[50_{7},175_2]_{(3)}$: {\bf Hoffmann-Singleton} &\\
\hline
$U_3(7)$& $(703,1075,49,141)$ & $[2107_{21},14749_3]_{(8)}$  & 0.991\\
\hline
$U_4(2)$& $S_4(3)$ in table 6 & table 6: {\bf GQ(2,4)}, {\bf GQ(3,3)}, etc & .\\
\hline
$U_4(3)$& $24,64,16,5)$ & $[112_{10},280_4]_{(3)}$: srg, $GQ(3,3^2)$  &\\
        & $30,90,24,10)$ & $[162_{280},15120_3]_{(3)}$: srg &\\
				  & $101,303,81,42)$ & $[567_{15},2835_3]_{(5)}$: NH$(2,14;(2,4))$\dag &\\
\hline
$U_4(4)$& $[87,165,17,29]$& $[325_{17},1105_5]_{(3)}$: srg, $GQ(4,4^2)$ &\\
\hline
$U_4(5)$& $[204,396,84,37]$& $[756_{26},3276_6]_{(3)}$: srg, $GQ(5,5^2)$& \\
\hline
$U_5(2)$& $[33,101,15,9]$& $[165_{9},297_5]_{(3)}$: srg, $GQ(4,8)$ & 0.950\\
& $[36,112,16,7]$& $[176_{40},1408_5]_{(3)}$: srg & 0.923\\
& $[61,153,27,29]$& $[297_{5},165_9]_{(3)}$: srg, $GQ(8,4)$ & 0.953\\
\hline
\hline
$U_6(2)$& $[96,416,62,50]$& $[672_{1408},157696_6]_{(3)}$: srg, $pg(11,15,3)?$ & \\
& $[99,437,63,48]$& $[693_{27},891_{21}]_{(3)}$: srg, $pg(20,8,5)?$& \\
& $[129,459,81,112]$& $[891_{21},6237_{3}]_{(4)}$: NH$(2,20;4)$\dag&\\
\hline
\end{tabular}
\caption{The non trivial configuration \lq stabilized' (bold) or \lq defined' by unitary groups with their corresponding $\mathcal{D}$ signature. (\dag) Groups $U_4(3)$ and $U_6(2)$ define two large near hexagons of order $(2,14)$ ans $(2,20)$, respectively: see \cite{deBruyn2006} for details about the notation.}
\end{center}
\end{table}

The unitary class of simple groups is a very rich one. It defines many generalized quadrangles, the hexagons $GH(2,2)$ associated to $3$-qubit contextuality (as shown in Sec. 2, table 3),  and two near hexagons including the largest of \lq slim dense' near hexagons on $891$ points, as shown in Table 8 \cite{deBruyn2006}. Whether such configurations have a physical relevance is unknown at the present time. Since unitary groups play a role as normalizers of Pauli groups, it may be expected that some of these geometries occur in the context of quantum error correction and Clifford groups \cite{Planat2010}.

Let us feature the $U_3(4)$ configuration. One defines the $3$-dimensional unitary space $U$ over the field $\mathbb{F}_{16}$, the projective space $\mathbb{P}(U)$ and a nondegenerate Hermitean form $(.,.)$ on $U$. The space $\mathbb{P}(U)$ consists of $65$ isotropic points $x$ satisfying $(x,x)=0$, $x \ne (0,0,0)$, and $208$ non-isotropic points satisfying $(x,x) \ne 0$. There exist $416$ orthogonal bases, that is, triples of mutually orthogonal non-isotropic points. The resulting configuration $[208_{6},416_3]_{(5)}$ has been shown to be related to a $3-(66,16,21)$ design used to construct the Suzuki sporadic group Suz  \cite{Brouwer2009} (see also table 12).

In passing, it is noticeable to feature the hyperplane structure of the $U_3(4)$ configuration. A basic hyperplane is defined from points of the collinearity graph that are either at minimum or maximal distance from a selected vertex. There are $208$ such hyperplanes. The other hyperplanes may be easily built from Velkamp sums $H \oplus H'$ of the basic hyperplanes $H$ and $H'$, where the set theoretical operation $\oplus$ means the complement of the symmetric difference $(H\cup H')\setminus (H \cap H')$ (as in \cite{Saniga2015}). One finds $10$ distinct classes of hyperplanes totalizing $2^{16}$ hyperplanes.

\subsection{Orthogonal}
\noindent

The geometries carried by orthogonal simple groups of small index are listed in Table 9. It is noticeable that some representations are associated to the non-intersection of maximum commuting sets for three qubits [from $O^+_8(2):2$)] and three qutrits [from $O_7(3)$ or $O^+_8(3)$]. These geometries are introduced in \cite[Table 2]{Pauli2011}. The srg's are identified in \cite{Brouwer}.  

\begin{figure}[ht]
\centering 
\includegraphics[width=8cm]{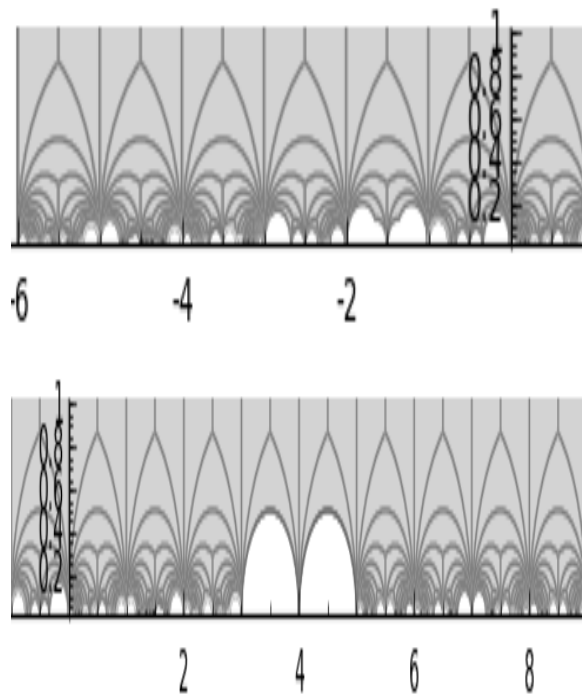}
\caption{A schematic of the hyperbolic polygon on $765$ tiles corresponding to the permutation group of $O_8^{-}(2)$. The picture is split into two horizontal parts.}
\label{Fig2}
\end{figure}

Several of the configurations arising from simple orthogonal groups are of type $\mathcal{G}_i$, for some $i\ge 1$. This includes the configurations attached to polar (strongly regular) graphs of $O^-_8(2)$ (on $119$ points), $O^-_8(3)$ (on $1066$ points) and $O^-_{10}(2)$ (on $495$ points). 

\small
\begin{table}[ht]
\begin{center}
\begin{tabular}{|l|c|r|c|r|}
\hline \hline
group & $\mathcal{D}$-signature &  geometry & $\kappa$\\
\hline
$O_7(3)$& $(51,239,27,18)$ & $[351_{567},28431_7]_{(3)}$, srg, $NO^-(7,3)$&\\
& $(28,58,24,238)$ & $[364_{40},1120_{13}]_{(3)}$, srg, $\mathcal{G}_4$, $W_5(3)$, $3$QT&\\
& $(54,252,30,22)$ & $[378_{3159},199017_{6}]_{(3)}$, srg, pg$(13,8,4)$?&\\
& $(156,540,84,151)$ & $[1080_{28431},3838185_{8}]_{(3)}$, srg, $NO^+(8,3)$&\\
& $(160,560,88,157)$ & $[1120_{13},3640_{4}]_{(4)}$, srg, DQ$(6,3)$, $3$QT$^*$&\\
\hline
$O_8^+(2):2$& $(12,92,12,3)$ & $[120_{28},1120_3]_{(3)}$, srg, $NO^+(8,2)$, pg$(7,8,4)$& 0.817\\
& $(15,99,11,6)$ & $[135_{64},960_9]_{(3)}$, srg, $\mathcal{G}_4$, pg$(8,7,4)$, $3$QB$^*$& 0.770\\
& $(96,624,72,85)$ & $[960_{36},4320_8]_{(4)}$&0.923\\
\hline
$O_8^-(2)$& $(41,63,7,5)$ & $[119_{45},765_7]_{(3)}$, $O_8^-(2)$ polar srg, $\mathcal{G}_3$, pg$(6,8,3)$?&\\
& $(46,72,8,6)$ & $[136_{135},2295_8]_{(3)}$, srg, NO$^-(8,2)$&\\
& $(267,389,45,33)$ & $[765_{7},1071_5]_{(4)}$, $\mathcal{G}_1$: NH$(4,6)$&\\
& $(552,832,96,77)$  & $[1632_{280},152320_3]_{(5)}$&\\
\hline
$O_8^+(3)$& $(216,604,84,89)$ &  as for $O_7(3)$, index $1080$ &\\
& $(224,616,88,97)$ &  as for $O_7(3)$, index $1120$ &\\
\hline
$O_8^-(3)$& $(274,598,26,85)$ & $[1066_{280},22960_{13}]_{(3)}$, $O_8^-(3)$ polar srg, $\mathcal{G}_4$, pg$(12,27,4)$? &\\
\hline
\hline
$O_{10}^+(2)$& $(38,376,16,34)$ &  index $496$, srg, NO$^+(10,2)$, pg$(15,16,8)$ &0.836\\
& $(45,391,15,39)$ &  index $527$, srg, pg$(16,15,8)$ & 0.759\\
& $(135,1335,117,355)$ &  index $2295$, rank $3$, $4$QB$^*$ & \\
\hline
$O_{10}^-(2)$& $(99,303,15,40)$ &  $[495_{765},25245_{15}]_{(3)}$, $O_{10}^-(2)$ polar srg , $\mathcal{G}_7$, pg$(14,16,7)$ &\\
& $(108,336,16,35)$ &  $[528_{2295},75735_{16}]_{(3)}$, srg, $\mathcal{G}_8$, NO$^-(10,2)$ &\\
\hline
\end{tabular}
\caption{The non trivial configuration \lq defined' by orthogonal groups with their corresponding $\mathcal{D}$ signature. The notation $3$QB$^*$ (resp. $3$QT$^*$ ) means that we are dealing with the geometry associated to the non-intersection of the maximum commuting sets built with the three-qubit (resp. three-qutrit) observables. Several configurations are of type $\mathcal{G}_i$. The near hexagon $O_8^{-}(2)$ on $765$ points is described in the text.}
\end{center}
\end{table}

\normalsize

\subsubsection*{The near hexagon $O_{8}^{-}(2)$}
\noindent

There exists a near polygon (thus of type $\mathcal{G}_1$) built from $O^-_8(2)$ (on $765$ points) that seems to have been unnoticed. The configuration is of the type $[765_{7},1071_5]_{(4)}$ with collinearity graph of spectrum $[28^1,11^{84},1^{476},-7^{204}]$ and diameter $3$ corresponding to a near hexagon of order $(4,6)$. Since the permutation representation is a subgroup of the modular group $\Gamma=PSL(2,\mathbb{Z})$, it is possible to see the dessin $\mathcal{D}$ as an hyperbolic polygon  $\mathcal{D}^{H}$. As in \cite{Planat2015,Moonshine2015},
the genus $g$ of $\mathbb{D}$ equals that of the hyperbolic polygon $\mathcal{D}^{H}$, a face of $\mathbb{D}$ corresponds to a cusp of $\mathcal{D}^{H}$, the number of black points (resp. of white points) of $\mathbb{D}$ is $B=f+\nu_3-1$ (resp. $W=n+2-2g-B-c$), where $f$ is the number of fractions,  $c$ is the number of cusps, $\nu_2$ and $\nu_3$ are the number of elliptic points of order two and three of $\mathcal{D}^{H}$, respectively. In the present case, the polygon $\mathcal{D}^{H}$ is associated to a non-congruence subgroup of level $17$ of $\Gamma$ and $(n,g,\nu_2,\nu_3,c,f)=(765,33,13,18,45,250)$. A schematic of $\mathcal{D}^{H}$ is shown in Fig. 2.

\subsection{Exceptional}
\noindent

A few exceptional groups of low index and low rank are defining well known generalized polygons GH$(2,2)$ and its dual, GH$(4,4)$ and its dual, GH$(2,8)$, the Ree-Tits octagon GO$(2,4)$, as well as two extra $\mathcal{G}_1$ geometries [coming from Sz$(8)$]. This is summarized in Table 10.

\small
\begin{table}[ht]
\begin{center}
\begin{tabular}{|l|c|r|c|r|}
\hline \hline
group & $\mathcal{D}$-signature &  geometry & $\kappa$\\
\hline
$G_2(2)'$ & $U_3(3)$ in table 8 & srg, $[63_3]$; GH$(2,2)$, GH$(2,2)$ dual & 0.846, 0.820\\
\hline
$G_2(4)$& $(88,224,32,37)$ & $[416_{8400},698830_5]_{(3)}$: srg, part of Suz graph&\\
& $(273,725,105,132)$ & $[1365_5]_{(4)}$: GH$(4,4)$&\\
& $(277,693,105,146)$ & $[1365_5]_{(4)}$: GH$(4,4)$ dual&\\
\hline
$^2F_4(2)'$ & $(585,923,135,57)$ & $[1755_{5},2925_3]_{(5)}$: GO$(2,4)$ (Ree-Tits)& 0.988 \\
\hline
Sz$(8)$ &(146,288,112,8])& $[560_{13},1820_4]_{(17)}$ & 0.971\\
 &(370,736,292,30) & $[1456_{5},1820_4]_{(79)}$ & 0.980 \\
\hline
$3D_4(2)$ &(95,419,63,122) & $[819_{9},2457_3]_{(4)}$: GH$(2,8)$&\\
\hline
\end{tabular}
\caption{Small non-trivial configurations \lq defined' by exceptional groups of Lie type. The most remarkable configurations are generalized hexagons, their duals and generalized octagon GO$(2,4)$.}
\end{center}
\end{table}
\normalsize

\subsection{Sporadic}
\noindent

Finally, small index representations of small sporadic groups lead to geometries of various types. The results are split into three tables: configurations arising from Mathieu groups in table 11, from Leech lattice groups in table 12 and the remaining ones -small sections of the Monster group and pariahs- in table 13. 
Niticeable geometries arising from sporadic groups are the $M_{24}$ near hexagon $NH(2,14)$ on $759$ points, the $J_2$ near octagon $NO(2,4)$ on $315$ points and Tits generalized octagon $GO(2,4)$ on $1755$ points. Another remarkable geometry is the one built from the McL graph on $275$ points, which is found to be of type $\mathcal{G}_2$, see also https://www.win.tue.nl/$\sim$aeb/graphs/McL.html for details about the McL graph.

\vspace*{.2cm}

This closes our investigation between simple groups and finite geometries. The contextuality parameter $\kappa$, when it is known, is the highest (exceeding $0.97$) for graphs associated to standard representations of $L_2(32)$, $U_3(4)$, $U_3(7)$, exceptional groups $2 F_4'(2)$ and $Sz(8)$, and sporadic groups such $M_{23}$, $M_{24}$, $Co_2$, $\mbox{McL}$, $He$, $Fi_{22}$, $T$, etc.

\begin{table}[ht]
\begin{center}
\begin{tabular}{|l|c|r|c|r|}
\hline \hline
group & $\mathcal{D}$-signature &  geometry & $\kappa$\\
\hline
$M_{11}$ & $(17,31,5,2)$& $[55_{9},165_3]_{(3)}$: srg, $T(11)$& \\
& $(20,38,6,2)$& $[66_{15},495_2]_{(4)}$ &\\
& $(45,89,15,9)$& $[165_{4},220_3]_{(8)}$ &\\
\hline
$M_{12}$ & $(22,38,6,1)$& srg, $[66_{10},220_3]_{(3)}$: $T(12)$ &\\
\hline
$M_{22}$ & $(25,45,7,1)$& srg, $[77_{16},616_2]_{(3)}$: srg, $S(3,6,22)$& 0.891\\
& $(48,96,16,9)$ & $[176_{210},9320_4]_{(3)}$: srg, $S(4,7,23)\setminus S(3,6,22)$& 0.953\\
& $(65,127,21,10)$ & $[231_{10},330_7]_{(4)}$: srg, $M_{22}$ graph \cite{Brouwer}& 0.955\\
& $(92,178,30,16)$& $[330_{77},1155_2]_{(5)}$ & 0.961\\
& $(162,320,56,40)$& $[616_{2},77_{16}]_{(5)}$& 0.973\\
\hline
$M_{23}$ & $(73,141,11,15)$&  $[253_{21},1771_3]_{(3)}$: srg, $T(23)$ & 0.971\\
& $(140,274,22,36)$&  $[506_{15},3795_2]_{(4)}$ & 0.984\\
& $(338,672,56,112)$&  $[1288_{165},106260_2]_{(4)}$ & 0.993\\
& $(469,931,77,148)$&  $[1771_{20},17710_2]_{(8)}$ & 0.992 \\
\hline
$M_{24}$ & $(102,144,12,10)$&  $[276_{22},2024_3]_{(3)}$: srg, $T(24)$ & 0.972\\
& $(267,387,33,37)$&  $[759_{15},3795_3]_{(4)}$: NH$(2,14;2)$& 0.990\\
& $(436,668,56,65$)& index $1288$: srg, pg$(22,35,14)?$ & 0.994\\
\hline
\end{tabular}
\caption{Small non-trivial configurations \lq defined' by Mathieu groups. A noticeable geometry is the $M_{24}$ near hexagon on $759$ points.}
\end{center}
\end{table}

\begin{table}[ht]
\begin{center}
\begin{tabular}{|l|c|r|c|r|}
\hline \hline
group & $\mathcal{D}$-signature &  geometry & $\kappa$\\
\hline
HS & $(20,60,10,6)$& $[100_{22},1100_2]_{(3)}$: srg & 0.903\\
& $(220,580,100,101)$& $[1100_{2},10_{22}]_{(5)}$&\\
\hline
$J_2$ & $(36,50,16,0)$& $[100_{336},8400_4]_{(3)}$: srg, Hall-Janko graph & 0.930\\
& $(196,146,40,0)$& $[280_{12},840_4]_{(4)}$: srg &\\
& $(105,165,45,1)$& $[315_{5},525_3]_{(6)}$: NO$(2,4)$ \cite{CohenTits1985}&\\
& $(179,265,75,4)$& $[525_{3},315_5]_{(6)}$&\\
& $(286,428,120,4)$& $[840_{5},1050_4]_{(7)}$&\\
& $(336,522,144,4)$& $[1008_{6},2016_3]_{(11)}$&\\
& $(604,910,258,15)$& $[1800_{70},42000_3]_{(18)}$&\\
\hline
Co$_2$ & $(460,1292,96,227)$&  srg$(2300,891,378,324)$\cite{Meixner1994}& 0.992\\
\hline
McL & $(55,155,25,21)$& $[275_{280},15400_5]_{(3)}$: srg, $\mathcal{G}_2$ \cite{McLaughin1969} & 0.974\\
& $(405,1065,185,186)$ & $[2025_{1155},779625_3]_{(4)}$&\\
\hline
Suz & $(594,912,138,70)$& $[1782_{1365},405405_6]_{(3)}$: srg \cite{Brouwer2009,Suzuki1969}& \\
\hline
\end{tabular}
\caption{Small non-trivial configurations \lq defined' by Leech lattice groups. Noticeable geometries are the $J_2$ near octagon on $315$ points and the Co$_2$ geometry that is locally the $U_6(2)$ near hexagon. Another remarkable configuration of type $\mathcal{G}_2$ is attached to the permutation representation on $275$ points of the McL group.}
\end{center}
\end{table}

\begin{table}[ht]
\begin{center}
\begin{tabular}{|l|c|r|c|r|}
\hline \hline
group & $\mathcal{D}$-signature &  geometry & $\kappa$\\
\hline
He & $(294,1106,122,269)$&  $[2058_{4896},3358656_{3}]_{(5)}$ & 0.984\\
\hline
Fi$_{22}$ & $(270,2102,320,410)$&  $[3510_{891},142155_{22}]_{(3)}$: srg & 0.997\\
Fi$_{23}$ & $(10575,16183,1163,1876)$&  srg, index $31671$ &\\
Fi$_{24}'$ & $(102312,155224,10584,19409)$&  srg, index $306936$&\\
\hline
$J_1$ & $(92,138,38,0)$& $[266_{11},1463_2]_{(5)}$: Livingstone graph&\\
& $(355,525,151,8)$& $[1045_{8},4180_2]_{(11)}$&\\
& $(491,747,209,9)$& $[1463_{6},2926_3]_{(22)}$&\\
& $(780,520,220,11)$& $[1540_{19},14630_2]_{(21)}$&\\
& $(532,804,228,17)$& $[1596_{11},8778_2]_{(19)}$&\\
\hline
Ru & $(1054,2030,316,331)$ & srg, index $4060$&\\
\hline
T=$^2F_4(2)'$ & $(585,923,135,57)$ & $[1755_{5},2925_3]_{(5)}$: GO$(2,4)$ \cite{Tits2002} & 0.988\\
 & $(774,1152,180,100)$ & $[2304_{26},14976_4]_{(7)}$& 0.988\\
\hline
\hline
\end{tabular}
\caption{Non-trivial configurations \lq defined' by small sections of the Monster group, the Pariah groups $J_1$ and $Ru$, and Tits group $T$.}
\end{center}
\end{table}

\section{Conclusion}
\noindent

We explored two-generator permutation representations of simple groups, as listed in the Atlas \cite{Atlasv3}, with the viewpoint of Grothendieck's dessins d'enfants and the finite geometries associated to them, as started in our earlier work. A strong motivation for this work is the understanding of commutation structures in quantum information and their contextuality \cite{Dessins2014}-\cite{Moonshine2015}, \cite{Levay2013, Planat2016}. A wealth of known and new point-line configurations $\mathcal{G}$, and as much as possible their contextuality parameter $\kappa$, are defined from permutation representations $\mathcal{P}$ and their corresponding dessin $\mathcal{D}$, using the methodology described in Sec. 2. It is intriguing that the concept of a near polygon, defined in Sec. 2.3, may be usefully expanded to that of a geometry of type $\mathcal{G}_i$ ($i>1$) to qualify some of the new  configurations we found.
Looking at unitary groups of table 8, one observes that most configurations we obtained are of the near polygon type (that is of type $\mathcal{G}_1$) or have a strongly regular collinearity graph. But we do not know how to unify both aspects. To some extent, orthogonal simple groups, as well as exceptional groups of Lie type, have this common feature (as shown in Tables 9 and 10, respectively). 

It is much more involved to recognize the regularities of geometries defined from (small) sporadic groups (see tables 11 to 13). Many sporadic groups (including the Monster) are subgroups of the modular group, or even of the Hurwitz group $G=\left\langle a,b|a^2=b^3=(ab)^7\right\rangle$ \cite {Wilson2001}. It is a challenging question to relate the symmetric genus of such structures to the (much smaller) genus of the corresponding dessin d'enfant (and modular polygon) \cite{Moonshine2015}. Our down-to-earth approach of understanding quantum commutation and contextuality from representations of some simple groups is of course far from the concept of a VOA (vertex operator algebra) which is related to string theory and generalized moonshine \cite{Gannon2005}. As final note, let us mention F.~J. Dyson again. {\it So far as we know, the physical universe would look and function just as it does whether or not the sporadic groups existed. But we should not be too sure that there is no connection $\cdots$ We have strong evidence that the creator of the universe loves symmetry, and if he loves symmetry, what lovelier symmetry could he find than the symmetry of the Monster?} \cite{Dyson1983}.

\section*{Acknowledgements} The interaction of the two authors was made possible thanks to grants ERGS 1-2013-5527178 and FRGS 11-2014-5524676. Part of the work was completed during the visit of the first author to the Universiti Putra Malaysia in the context of SEAMS School: Algebra and their applications, Nov. 3 to 10 (2015).

\end{document}